\documentclass[twocolumn,showpacs,preprintnumbers,amsmath,amssymb]{revtex4}
\usepackage{graphicx}
\usepackage{dcolumn}
\usepackage{bm}
\usepackage{epsfig}

\begin{document}

\title{JLab Measurements of the $^3$He Form Factors
at Large Momentum Transfers}

\author
{
{A.~Camsonne}$^{22}$,
{A.T.~Katramatou}$^{11}$,
{M.~Olson}$^{20}$,
{A.~Acha}$^{4}$,
{K.~Allada}$^{12}$,
{B.D.~Anderson}$^{11}$,
{J.~Arrington}$^{1}$,
{A.~Baldwin}$^{11}$,
{J.-P.~Chen}$^{22}$,
{S.~Choi}$^{18}$,
{E.~Chudakov}$^{22}$,
{E.~Cisbani}$^{8,10}$,
{B.~Craver}$^{23}$,
{P.~Decowski}$^{19}$,
{C.~Dutta}$^{12}$,
{E.~Folts}$^{22}$,
{S.~Frullani}$^{8,10}$,
{F.~Garibaldi}$^{8,10}$,
{R.~Gilman}$^{16,22}$,
{J.~Gomez}$^{22}$,
{B.~Hahn}$^{24}$,
{J.-O.~Hansen}$^{22}$,
{D.~W.~Higinbotham}$^{22}$,
{T.~Holmstrom}$^{13}$,
{J.~Huang}$^{14}$,
{M.~Iodice}$^{9}$,
{X.~Jiang}$^{16}$,
{A.~Kelleher}$^{24}$,
{E.~Khrosinkova}$^{11}$,
{A.~Kievsky}$^{7}$,
{E.~Kuchina}$^{16}$,
{G.~Kumbartzki}$^{16}$,
{B.~Lee}$^{18}$,
{J.J.~LeRose}$^{22}$,
{R.A.~Lindgren}$^{23}$,
{G.~Lott}$^{22}$,
{H.~Lu}$^{17}$,
{L.E.~Marcucci}$^{7,15}$,
{D.J.~Margaziotis}$^{2}$,
{P.~Markowitz}$^{4}$,
{S.~Marrone}$^{6}$,
{D.~Meekins}$^{22}$,
{Z.-E.~Meziani}$^{21}$,
{R.~Michaels}$^{22}$,
{B.~Moffit}$^{14}$,
{B.~Norum}$^{23}$,
{G.G.~Petratos}$^{11}$,
{A.~Puckett}$^{14}$,
{X.~Qian}$^{3}$,
{O.~Rondon}$^{23}$,
{A.~Saha}$^{22}$,
{B.~Sawatzky}$^{21}$,
{J.~Segal}$^{22}$,
{M.~Shabestari}$^{23}$,
{A.~Shahinyan}$^{25}$,
{P.~Solvignon}$^{1}$,
{N.~Sparveris}$^{11,21}$,
{R.R.~Subedi}$^{23}$,
{R.~Suleiman}$^{22}$,
{V.~Sulkosky}$^{22}$,
{G.M.~Urciuoli}$^{8}$,
{M.~Viviani}$^{7}$,
{Y.~Wang}$^{5}$,
{B.B.~Wojtsekhowski}$^{22}$,
{X.~Yan}$^{18}$,
{H.~Yao}$^{21}$,
{W.-M.~Zhang}$^{11}$,
{X.~Zheng}$^{23}$,
{and L.~Zhu}$^{5}$
}


\affiliation{$^{1}$Argonne National Laboratory, Argonne, IL 60439, USA}
\affiliation{$^{2}$California State University, Los Angeles, CA 90032, USA }
\affiliation{$^{3}$Duke University (TUNL), Durham, NC 27708, USA}
\affiliation{$^{4}$Florida International University, Miami, FL 33199, USA}
\affiliation{$^{5}$University of Illinois at Urbana Champagne, Urbana, IL 61801, USA}
\affiliation{$^{6}$Istituto Nazionale di Fisica Nucleare, Sezione di Bari and University of Bari, 70126 Bari, Italy}
\affiliation{$^{7}$Istituto Nazionale di Fisica Nucleare, Sezione di Pisa, 56127 Pisa, Italy}
\affiliation{$^{8}$Istituto Nazionale di Fisica Nucleare, Sezione di Roma, 00185 Rome, Italy}
\affiliation{$^{9}$Istituto Nazionale di Fisica Nucleare, Sezione di Roma Tre, 00146 Rome, Italy}
\affiliation{$^{10}$Istituto Superiore di Sanit\`{a}, 00161 Rome, Italy}
\affiliation{$^{11}$Kent State University, Kent OH 44242, USA}
\affiliation{$^{12}$University of Kentucky, Lexington, KY 40506, USA}
\affiliation{$^{13}$Longwood University, Farmville, VA 23909, USA}
\affiliation{$^{14}$Massachusetts Institute of Technology, Cambridge, MA 02139, USA}
\affiliation{$^{15}$University of Pisa, 56127 Pisa, Italy}
\affiliation{$^{16}$Rutgers, The State University of New Jersey, Piscataway, NJ 08855, USA}
\affiliation{$^{17}$University of Science and Technology of China, Hefei, Anhui, 230026 P.R. China}
\affiliation{$^{18}$Seoul National University, Seoul 151-747, Korea}
\affiliation{$^{19}$Smith College, Northampton, MA 01063, USA}
\affiliation{$^{20}$St.~Norbert College, De Pere WI 54115, USA}
\affiliation{$^{21}$Temple University, Philadelphia, PA 19122, USA}
\affiliation{$^{22}$Thomas Jefferson National Accelerator Facility, Newport News, VA 23606, USA}
\affiliation{$^{23}$University of Virginia, Charlottesville, VA 22904, USA}
\affiliation{$^{24}$College of William and Mary, Williamsburg, VA 23185, USA}
\affiliation{$^{25}$Yerevan Physics Institute, Yerevan 375036, Armenia}

\centerline{The Jefferson Lab Hall A Collaboration}

\centerline {Dated: October 17, 2016}

\begin{abstract}

The charge and magnetic form factors, $F_C$ and $F_M$, of $^3$He have been extracted
in the kinematic range 25 fm$^{-2}$ $\le Q^2 \le 61$ fm$^{-2}$ from elastic electron
scattering by detecting $^3$He recoil nuclei and electrons in coincidence with the
High Resolution Spectrometers of the Hall A Facility at Jefferson Lab.
The measurements are indicative of a second diffraction minimum for the magnetic
form factor, which was predicted in the $Q^2$ range of this experiment, and of a
continuing diffractive structure for the charge form factor.
The data are in qualitative agreement with theoretical calculations based on
realistic interactions and accurate methods to solve the three-body nuclear problem.

\vspace* {-.1in}

\end{abstract}

\pacs{25.30.Bf, 13.40.Gp, 27.10.+h, 24.85.+p}

\maketitle

Elastic electron scattering from nuclei has been a basic tool in the study
of their size and associated charge and magnetization distributions \cite{ub71}.
It allows for the extraction of their electromagnetic (EM) form factors,
which in the case of few-body nuclear systems can be compared with state-of-the-art
theoretical calculations.  The few-body form factors are considered the
``observables of  choice''~\cite{ma98} for testing the nucleon-meson standard model
of the nuclear  interaction and the associated EM current operator~\cite{ca98}.
They provide fundamental information on the internal structure and dynamics of
light nuclei as they are, at the simplest level, convolutions of the nuclear ground
state wave function with the EM form factors of the constituent nucleons.
The theoretical calculations for these few-body observables are very
sensitive to the model used for the nuclear EM current operator, especially
its meson-exchange-current (MEC) contributions.  Relativistic corrections
and possible admixtures of multi-quark states in the nuclear wave function
might also be relevant~\cite{ca98}.  Additionally, at large momentum
transfers, these EM form factors may offer a unique opportunity to uncover
a possible transition in the description of elastic electron scattering off
of few-body nuclear systems, from meson-nucleon to quark-gluon degrees of
freedom, as predicted by the dimensional-scaling
quark model~(DSQM) and perburbative QCD~(pQCD)~\cite{br76,br84}.

Experimentally, the few-body form factors are determined
from elastic electron-nucleus scattering using high intensity
beams, high density targets, and large solid angle magnetic spectrometers.
There have been extensive experimental investigations
of the few-body form factors over the past 50 years at almost every electron
accelerator laboratory~\cite{si01,gi02}, complemented by equally extensive
theoretical calculations and predictions~\cite{ca98,gi02,ho12}.  A recent
review can be found in Ref.~\cite{ma16}.

This work focuses on a measurement of the $^3$He EM form factors at
Jefferson Lab (JLab).
The cross section for elastic scattering of a relativistic electron from the
spin 1/2 $^3$He nucleus is given, in the one-photon exchange approximation
and in natural units,
by the formula~\cite{ha85}:
\begin{equation}
{ {d\sigma} \over {d\Omega} } =
{ \left( {d\sigma} \over {d\Omega} \right)_{NS} } 
 {\left[ A(Q^2)+B(Q^2)\tan^2 \left( {\theta \over 2} \right) \right] },
\label{roseform}
\end{equation}
where
\begin{equation}
{ \left( {d\sigma} \over {d\Omega} \right)_{NS} } =
{ {(Z\alpha)^2 E^\prime \cos^2 \left( {\theta \over 2} \right)}
\over {4 E^3  \sin^4 \left( {\theta \over 2}\right)} }
\label{nostructure}
\end{equation}
is the cross section for the scattering of a relativistic electron by
a structureless nucleus, and $A$ and $B$ are the elastic structure
functions of $^3$He:
\begin{equation}
{   A(Q^2) = {   { F^2_C(Q^2) +
{\mu}^2 \tau F^2_M(Q^2) } \over {1 + \tau} }    },
\label{}
\end{equation}
\begin{equation}
{ B(Q^2) =  2 \tau {\mu}^2 F^2_M(Q^2) },
\label{}
\end{equation}
with $F_C$ and $F_M$ being the charge and magnetic form factors of the nucleus.
Here, $\alpha$ is the fine-structure constant, $Z$ and $\mu$ are the nuclear
charge and magnetic moment,
$E$ and $E'$ are the incident and scattered electron energies,
$\theta$ is the electron scattering angle, $Q^2 = 4 E E' \sin^2 (\theta/2)$
is minus the four-momentum transfer squared, and $\tau=Q^2/4M^2$ with $M$ being the
nuclear mass.

The three-body form factors have been theoretically investigated by several groups,
using different techniques to solve
for the nuclear ground states, and a variety of models for the nuclear
EM current~\cite{ha86,st87,sc90,wi91}.  The most recent calculation of
the $^3$H and $^3$He form factors in the $Q^2$-range of the experiment
is that of Refs.~\cite{ma98,ma05}.  It uses the pair-correlated
hyperspherical harmonics (HH) method~\cite{ki08} to construct high-precision
nuclear wave functions and goes beyond the impulse approximation (IA)
(where the electron interacts with just one of the nucleon constituents)
by including MEC, whose main contributions are constructed to satisfy
the current conservation relation with the given Hamiltonian~\cite{ma05}.
Part of the present work is the extension of the above method to evolve the $^3$He
$F_C$ and $F_M$ form factors (see Figures~1-3) to large momentum transfers,
using the (uncorrelated) HH expansion to solve for the $^3$He wave function
from the Argonne AV18
nucleon-nucleon ($NN$) and
Urbana UIX
three-nucleon ($3N$) interactions~\cite{wi95}.  The calculations
include MEC contributions arising from $\pi$-, $\rho$-
and $\omega$-meson exchanges, as well as the $\rho\pi\gamma$ and
$\omega\pi\gamma$ charge transition couplings.

At large $Q^2$, elastic scattering from few-body nuclear systems like $^3$He
may be partly due to, or even dominated by, contributions from the electron's
interaction with the nucleons' constituent quarks.
A purely phenomenological ``hybrid quark-hadron'' approach includes
multi-quark states for overlapping nucleons in the nuclear
wave function, which augment the IA approach~\cite{di90}.
The field theory approach of the DSQM, later substantiated within the
pQCD framework~\cite{br84}, is based on dimensional scaling of high energy
amplitudes using quark counting. This leads to the asymptotic prediction
$\sqrt{F_C(Q^2)} \sim (Q^2)^{1-3A}$, where $A$~=~3 for the $^3$He case
(see Ref. \cite{br76}).

The experiment (E04-018) used the Continuous Electron Beam (100$\%$ duty factor)
Accelerator and Hall A Facilities of JLab.
Electrons scattered from a high density cryogenic $^3$He target were
detected in the Left High Resolution Spectrometer (e-HRS).  To
suppress backgrounds and unambiguously separate elastic from inelastic
processes, recoil helium nuclei were detected in the Right HRS (h-HRS)
in coincidence with the scattered electrons.
The incident-beam energy ranged between 0.688 and 3.304 GeV.
The beam current ranged between 29 and 99$\mu$A.
The cryogenic target system contained gaseous $^3$He and liquid hydrogen
cells of length $T$=20~cm.  The $^3$He gas was pressurized to 13.7-14.2~atm at
a temperature of 7.1-8.7~K, resulting in a density of 0.057-0.070 g/cm$^3$.
Two Al foils separated by 20~cm
were used to measure any possible contribution to the cross section from the
Al end-caps of the target cells.

\begin{figure} [t]
\includegraphics[width=69mm, angle=90]{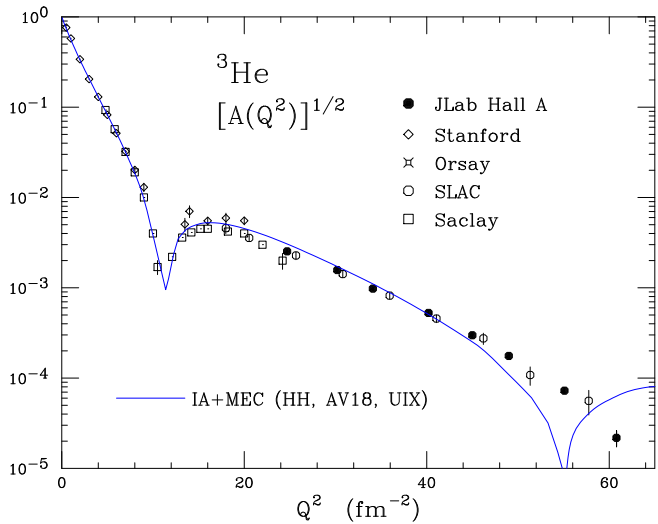}
\caption{\label{fig:a} $^3$He elastic structure function $A(Q^2)$ data
from this experiment, compared to selected previous data and
the present theoretical calculation with the hyperspherical harmonics variational
method (see text).}
\vspace* {-.15in}
\end{figure}

Scattered electrons were detected in the e-HRS using two planes of
scintillators to form an ``electron'' trigger, a pair of drift
chambers for electron track reconstruction, and a gas threshold
\v{C}erenkov counter and a lead-glass calorimeter for electron
identification.  Recoil helium nuclei were detected in the h-HRS
using two planes of scintillators to form a ``recoil'' trigger
and a pair of drift chambers for recoil track reconstruction.
The event trigger consisted of a coincidence between the two HRS triggers.
Details on the Hall A Facility and all associated
instrumentation used are given in Refs.~\cite{ca14,al04}.

\begin{figure} [t]
\vspace* {.25in}
\includegraphics[width=69mm, angle=90]{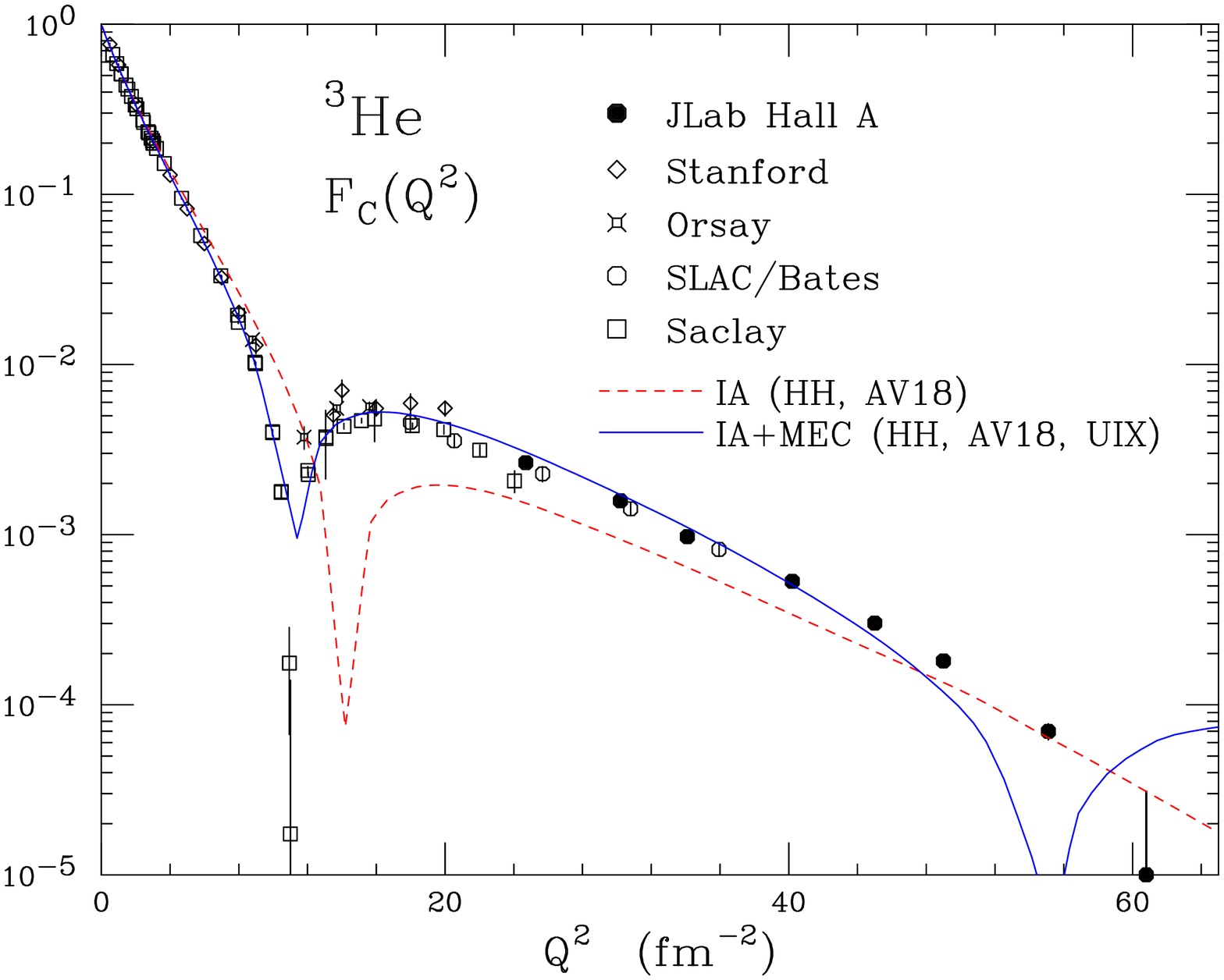}
\caption{\label{fig:fc} $^3$He charge form factor $F_C$ data
from this experiment, compared to selected previous data and
the present theoretical calculation with the hyperspherical harmonics
variational method (see text).}
\vspace* {-.15in}
\end{figure}

Particles in the e-HRS were identified as electrons on the basis of
a minimal pulse height in the \v{C}erenkov counter
and the energy deposited in the calorimeter, consistent with the
momentum as determined from the drift chamber track using the spectrometer's
optical properties.
Particles in the h-HRS were identified as $^3$He nuclei on the basis of their
energy deposition in the first scintillator plane.
Electron-$^3$He~($e$-$^3$He) coincidence events, consistent with elastic
kinematics,
were identified using the relative time-of-flight (TOF) between
the electron and recoil triggers after imposing the above particle
identification ``cuts".
To check the overall normalization, elastic
$e$-proton ($e$-$p$) scattering was measured at several kinematics.
The $e$-$p$ data are in excellent agreement with the world data, as described
in Ref.~\cite{ca14}.

\begin{figure} [t]
\includegraphics[width=69mm, angle=90]{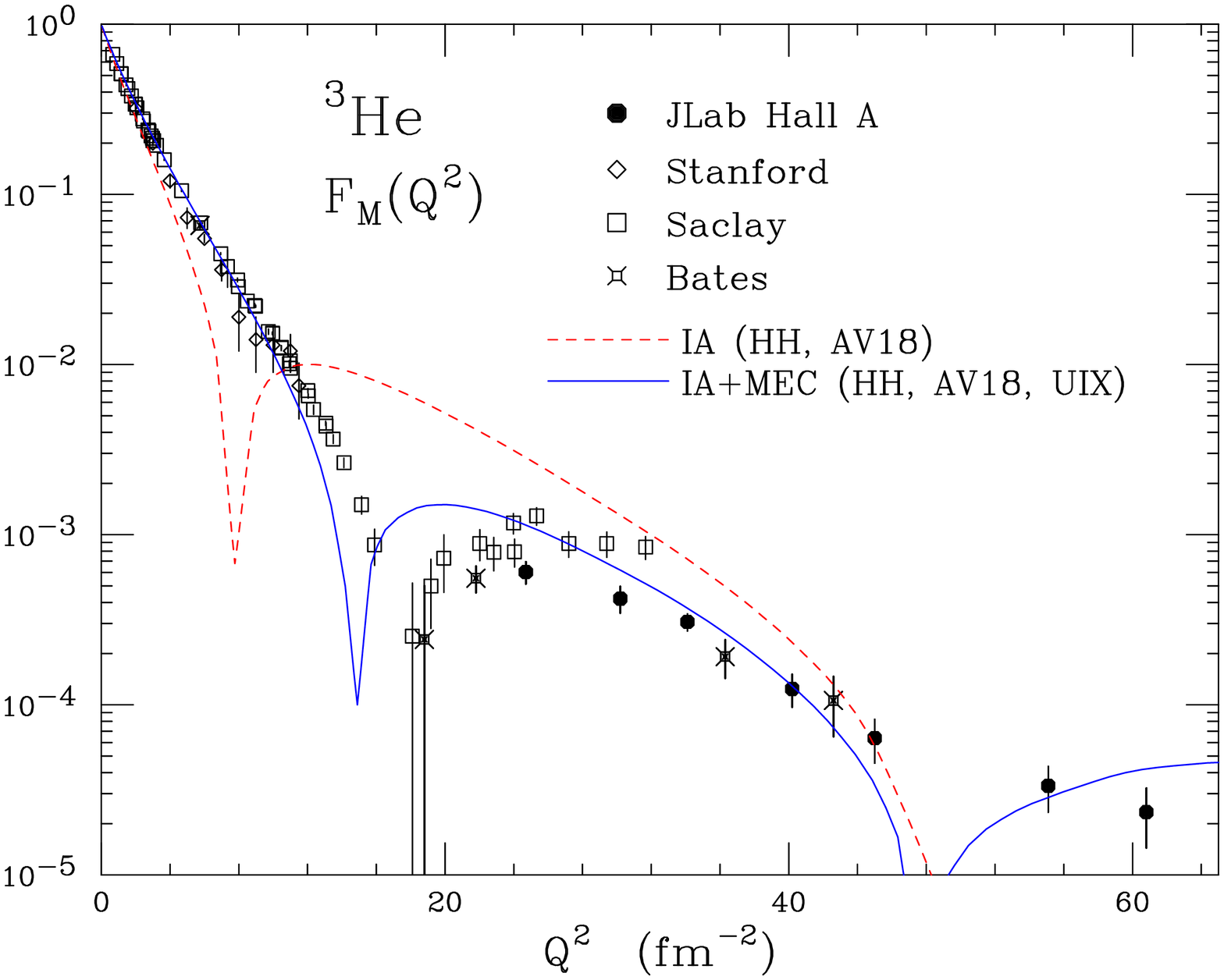}
\caption{\label{fig:fm} $^3$He magnetic form factor $F_M$ data
from this experiment, compared to selected previous data and
the present theoretical calculation with the hyperspherical harmonics
variational method (see text).}
\vspace* {-.15in}
\end{figure}

The elastic $e$-$^3$He cross section values were calculated using the formula:
\begin{equation}
\left[ \frac {d\sigma} {d\Omega} (E,\theta) \right]_{exp}=\frac {N_{er}C_{cor}}
{N_b N_t (\Delta\Omega)_{MC} F(Q^2,T)},
\end{equation}
where $N_{er}$ is the number of electron-recoil $^3$He elastic events,
$N_b$ is the number of incident beam electrons, $N_t$ is the
number of target nuclei/cm$^2$, $(\Delta\Omega)_{MC}$ is the
effective coincidence solid angle (which includes most radiative effects)
from a Monte Carlo simulation,
$F$ is the portion of the radiative corrections that depends only on $Q^2$
and $T$ (1.07-1.10)~\cite{moca},
and $C_{cor}=C_{det}C_{cdt}C_{rni}C_{den}$.  Here, $C_{det}$ is the
correction for the inefficiency of the \v{C}erenkov counter
and the calorimeter (1.01) (the scintillator counter hodoscopes
were found to be essentially 100$\%$ efficient),
$C_{cdt}$ is the computer dead-time correction (1.04-1.56),
$C_{rni}$ is a correction for losses of recoil nuclei due to
nuclear interactions in the target cell and vacuum windows (1.02-1.08),
and $C_{den}$ is a correction to
the target density due to beam heating effects
(ranging between 1.02 at 29$\mu$A and 1.07 at 99$\mu$A).
There were no contributions to the elastic $e$-$^3$He cross section from events
originating in the target cell end-caps, as determined from runs with the
empty replica target.  The $e$-$p$ elastic cross section values were
determined similarly.

The effective coincidence solid angle was evaluated with a
Monte Carlo computer code that simulated elastic
electron-nucleus scattering under identical conditions as our
measurements~\cite{moca}.  The code tracked scattered electrons and
recoil nuclei from the target to the detectors through the
two HRS systems using optical models based on magnetic field
measurements and precision position surveys of their elements.
The effects from ionization energy losses
and multiple scattering in the target and vacuum windows were
taken into account for both electrons and recoil nuclei.
Bremsstrahlung radiation losses
for both incident and scattered electrons in the target and
vacuum windows, as well as internal radiative effects, were also
taken into account.
It should be noted that the two-photon exchange effect is
not included in the radiative corrections implementation.  A credible
correction to the data for this effect should be based on established
complemantary calculations, which are not yet fully available for the
entire kinematic range of our measurements.  A correction (not only for
this, but for all available $^3$He elastic data sets) will have
to wait for the completion and further understanding of ongoing
calculations~\cite{2gam}.

\begin{table}
\begin{tabular}{cccc}
\hline
$E$ & $\theta$ & $Q^2$     & $d\sigma/d\Omega$ \\
GeV & deg.     & fm$^{-2}$ & cm$^2$/sr         \\
\hline
3.304  & 17.52  &  24.7 & $(2.29\pm0.12)\times10^{-35}$ \\
0.7391 & 97.78  &  24.7 & $(2.80\pm0.20)\times10^{-37}$ \\
3.304  & 19.50  &  30.2 & $(5.16\pm0.29)\times10^{-36}$ \\
0.7391 & 118.99 &  30.2 & $(3.95\pm0.38)\times10^{-38}$ \\
0.6876 & 139.99 &  30.2 & $(1.51\pm0.19)\times10^{-38}$ \\
3.304  & 20.83  &  34.1 & $(1.57\pm0.10)\times10^{-36}$ \\
0.8157 & 113.01 &  34.1 & $(1.43\pm0.13)\times10^{-38}$ \\
0.7394 & 139.91 &  34.1 & $(6.41\pm0.59)\times10^{-39}$ \\
3.304  & 22.82  &  40.2 & $(3.03\pm0.21)\times10^{-37}$ \\
0.8177 & 139.53 &  40.2 & $(1.24\pm0.16)\times10^{-39}$ \\
3.304  & 24.28  &  45.0 & $(7.56\pm0.75)\times10^{-38}$ \\
0.9330 & 119.94 &  45.0 & $(6.84\pm1.10)\times10^{-40}$ \\
0.8726 & 140.66 &  45.0 & $(3.24\pm0.51)\times10^{-40}$ \\
3.304  & 25.47  &  49.0 & $(2.13\pm0.35)\times10^{-38}$ \\
3.304  & 27.24  &  55.1 & $(2.77\pm0.39)\times10^{-39}$ \\
0.9893 & 140.31 &  55.1 & $(3.27\pm0.13)\times10^{-41}$ \\
3.304  & 28.86  &  60.8 & $(2.14\pm0.72)\times10^{-40}$ \\
1.052  & 140.51 &  60.8 & $(1.13\pm0.80)\times10^{-41}$ \\
\hline
\end{tabular}
\caption
{
Values of beam energy, scattering angle, effective $Q^2$, and elastic $e$-$^3$He
cross section with total error (statistical and systematic added in quadrature).
}
\label{dat1}
\vspace* {-0.05in}
\end{table}

\begin{table}
\begin{tabular}{ccc}
\hline
$Q^2$           & $|F_C|$   & $|F_M|$ \\
 fm$^{-2}$     &           &         \\
\hline
24.7 & $(2.65\pm0.06)\times10^{-3}$ & $(6.03\pm0.91)\times10^{-4}$  \\
30.2 & $(1.58\pm0.05)\times10^{-3}$ & $(4.21\pm0.75)\times10^{-4}$  \\
34.1 & $(9.73\pm0.34)\times10^{-4}$ & $(3.07\pm0.35)\times10^{-4}$  \\
40.2 & $(5.32\pm0.21)\times10^{-4}$ & $(1.24\pm0.27)\times10^{-4}$  \\
45.0 & $(3.02\pm0.16)\times10^{-4}$ & $(6.37\pm1.83)\times10^{-5}$  \\
49.0 & $(1.81\pm0.15)\times10^{-4}$ &        $-$                    \\
55.1 & $(6.97\pm0.72)\times10^{-5}$ & $(3.34\pm1.00)\times10^{-5}$  \\
60.8 & $(1.00\pm2.10)\times10^{-5}$ & $(2.34\pm0.90)\times10^{-5}$  \\
\hline
\end{tabular}
\caption{Effective $Q^2$, and $^3$He charge and magnetic form factors (absolute
values) with total errors (statistical and systematic added in quadrature).}
\label{dat2}
\vspace* {-0.2in}
\end{table}
 
The Rosenbluth cross section formula (1) is based on the assumption that the wave
functions of the incident and scattered electrons are described by plane waves.
In reality, the charge of the nucleus distorts these wave functions, necessitating
a correction to the formula~\cite{ub71}.  This Coulomb effect shifts the
$Q^2$ value of the interaction to an ``effective" value, given by
${Q^2}_{eff}=(1+3 Z \alpha \hbar c/2 R_{eq} E)^2 Q^2$, where $R_{eq}$ is the hard sphere
equivalent radius of the nucleus, $\hbar$ is the Planck constant and $c$ is the speed of light. 
This correction allows for a form factor extraction using a Rosenbluth separation of
cross section values determined, at each kinematic point, at the same $Q^2_{eff}$~\cite{hepl}.
This approach was followed in this experiment and the results are given in
terms of the effective $Q^2$ in Tables I and II and plotted in Figs 1-3.

At each kinematic point, the
``reduced'' cross section, $(d\sigma / d\Omega)_r$, defined using equations (1-4)
and the experimentally determined cross section $[d\sigma / d\Omega]_{exp}$
\begin{equation}
{ \left( {d\sigma} \over {d\Omega} \right)_r } =
{ \left( {d\sigma} \over {d\Omega} \right)_{exp} } 
{ \left( {d\sigma} \over {d\Omega} \right)_{NS}^{-1} } {(1+\tau)} =
{ \left( F_C^2 + {\mu}^2 {\tau \over \epsilon} F_M^2 \right) }
\label{reduced}
\end{equation}
was plotted, at same values of $Q^2_{eff}$, versus $\mu^2\tau/\epsilon$ (Rosenbluth plot),
and the $^3$He $F_C^2$ and $F_M^2$ values were extracted by a linear fit.  Here,
$\epsilon=[1+2(1+\tau) \tan^2(\theta/2)]^{-1}$ is the degree
of the longitudinal polarization of the exchanged virtual photon.
It should be noted that at $Q^2=49$~fm$^{-2}$ data were taken only at a forward angle
(25.47$^\circ$) and that the $F_C$ value was extracted under the resonable assumption that
the $F_M$ does not contribute to the cross section.

The $A(Q^2)$ values from this experiment
are shown in Fig.~\ref{fig:a} along with previous data from a SLAC 
experiment~\cite{ar78}, which performed elastic scattering at a fixed angle
$\theta=8^\circ$, and selected data from other laboratories~\cite{hepl,be72,am94}.
It is evident that the JLab and SLAC data sets are in excellent agreement.
Also shown is the present IA+MEC theoretical calculation (see below).
The (absolute) values of the $^3$He $F_C$ and $F_M$ from this work are 
shown in Figs.~\ref{fig:fc} and \ref{fig:fm} along with
previous Stanford~\cite{hepl}, Orsay~\cite{be72}, SLAC~\cite{ar78}, Saclay~\cite{am94} 
and MIT/Bates~\cite{na01} data.  Not shown, for clarity, are the low $Q^2$
MIT/Bates data~\cite{miba}.  In all three figures, the error bars represent statistical
and systematic uncertainties added in quadrature.
The new $F_C$ data are in excellent agreement with data
extracted from a Rosenbluth separation between SLAC forward
angle ($\theta=8^\circ$) cross sections and interpolations of backward angle
($160^{\circ}$) MIT/Bates~\cite{na01} cross sections, labeled as ``SLAC/Bates" data
in Fig. 2.
The new $F_M$ data are in excellent agreement with the MIT/Bates
data taken at $\theta=160^\circ$, but in very strong disagreement with the Saclay
data taken at $\theta=155^\circ$.
The $F_M$ datum at $Q^2$ = 24.7 fm$^{-2}$ has
been extracted from a Rosenbluth separation of a forward- and a medium-$\theta$ JLab-measured
cross section and an interpolated cross section from the 
$\theta=160^\circ$ MIT/Bates data set~\cite{na01}.  

An updated extension of the latest theoretical
calculation based on the IA with inclusion of MEC, which used the 
HH variational method to calculate the $^3$He wave function, as described above
and outlined 
in Ref. \cite{ma05}, was performed for this work and is shown in
Figs.~\ref{fig:fc} and \ref{fig:fm}.  The calculation is, in general,
in qualitative agreement with the data even at 
large momentum transfers where theoretical unceratinties may become
sizable.  Of note is the long-standing disagreement between the
calculation and the data in the $Q^2$ range around the first diffraction
minimum of the $^3$He $F_M$.
It is not presently clear if this is due to a missing piece of
important physics in the non-relativistic theory or to the need for
a fully relativistic calculation.  The presently available relativistic
calculation based on the Gross equation~\cite{gros} will be able to be compared
to the new data when the not-yet-calculated $\rho\pi\gamma$ interaction
current is included in this so called ``relativistic impulse approximation"
approach~\cite{ma16}.

It should be noted that all seminal, older calculations of the $^3$He
form factors (not shown in Figs.~\ref{fig:fc} and \ref{fig:fm}) based
on the Faddeev formalism~\cite{ha86,st87}
or the Monte Carlo variational method~\cite{sc90,wi91}, are in qualitative agreement
with the data in predicting a diffractive structure for both form factors, and
also indicative, in general, of large MEC contributions.
Also, it is evident that the diffractive pattern of the JLab data is incompatible
with the asymptotic-falloff DSQM prediction~\cite{br76}, and that
it supports the conclusion of
Ref.~\cite{ho90} that the onset of asymptotic scaling must be at a $Q^2$ value
much greater than 100 fm$^{-2}$, not presently accessible
at JLab for $^3$He.

In summary, we have measured the $^3$He charge and magnetic form factors
in the range 25 fm$^{-2}$ $\le Q^2 \le 61$ fm$^{-2}$.
The results are in qualitative
agreement with theoretical calculations based on the IA
with inclusion of MEC.
The new data strongly indicate the presence of an apparent second diffraction minimum
for the magnetic form factor in the vicinity of $Q^2$ = 50 fm$^{-2}$ as well as the possible
presence of a second diffraction minimum for the charge form factor located at
a $Q^2$ value just beyond 60 fm$^{-2}$.  The results will constrain
inherent uncertainties of the theoretical calculations and lead,
together with previous large $Q^2$ data on the deuteron~\cite{al99}, tritium
~\cite{am94} and $^4$He~\cite{ca14} EM form factors, to the
development of a consistent hadronic model describing the
internal EM structure and dynamics of few-body nuclear systems.

We acknowledge the outstanding support of the staff
of the Accelerator and Physics Divisions of JLab
that made this experiment possible.  We are grateful to
Drs. D.~Riska, R. Schiavilla, and R.~Wiringa
for kindly providing their theoretically motivating calculations for the
proposal of this experiment, and to Drs. F.~Gross,
R.~Schiavilla, and W.~Melnitchouk for valuable discussions and support.
This material is based upon work supported by the U.S. Department of Energy (DOE), Office of Science,
Office of Nuclear Physics under contract DE-AC05-060R23177.  This work was also supported by DOE
awards DE-AC02-06CH11357 and DE-FG02-96ER40950, National Science Foundation awards NSF-PHY-0701679,
NSF-PHY-1405814 and NSF-PHY-0652713, the Kent State University Council, and the INFN.

{}

\end{document}